# OVERVIEW AND PROSPECTS OF USING INTEGER SURROGATE KEYS FOR DATA WAREHOUSE PERFORMANCE OPTIMIZATION

Sviatoslav Stumpf and Vladislav Povyshev

ITMO University, Saint Petersburg, RUSSIA

## ABSTRACT

*The aim of this paper is to examine and demonstrate how integer-based datetime labels (integer surrogate keys for time) can optimize data-warehouse and time-series performance, proposing practical formats and algorithms and validating their efficiency on real-world workloads. It is shown that replacing standard DATE and TIMESTAMP types with 32- and 64-bit integer formats reduces storage requirements by 30–60% and speeds up query execution by 25–40%. The paper presents indexing, aggregation, compression, and batching algorithms demonstrating up to an eightfold increase in throughput. Practical examples from finance, telecommunications, IoT, and scientific research confirm the efficiency and versatility of the proposed approach.*

## 1. PROBLEM OVERVIEW

### 1.1. Relevance of the Problem

Objective: the study focuses on replacing native DATE/TIMESTAMP types with 32-bit (YYYYMMDD) and 64-bit integer encodings, and on the accompanying methods – indexing, range search, aggregation/binning, batching, compression, and TAI/UTC handling with example implementations in PostgreSQL and ClickHouse and use cases from finance, telecom, IoT, and scientific data.

Main subject of study: time series underpin many modern information systems: from financial analytics platforms and IoT devices to industrial monitoring and telemetry. Efficient storage and processing of such data is critical, as query speed, resource consumption, and system scalability depend on it. Modern technologies generate unprecedented volumes of temporal data. According to study [1], by 2028 the global data volume will reach 394 zettabytes, with up to 30% consisting of time-series data from IoT devices, financial systems, and telemetry. This explosive growth creates serious challenges for storage and processing systems.

Traditional databases often prove unprepared for such loads. As noted in [2], built-in time data types (TIMESTAMP, DATETIME) require significant resources. For example, storing billions of records with nanosecond precision in standard formats can take 30–50% more space compared to optimized solutions [3].

These problems are especially acute in real-world scenarios. In high-frequency trading (HFT), for example, nanosecond precision is required, as shown in [4]. Industrial IoT, according to [5], generates up to 15 TB of data daily with millisecond-level processing latency requirements. Telecom operators, per [6], process billions of temporal events every day (such as call records, communication-infrastructure events, and billing records). As TimescaleDB experts note [7], standard time formats in practical applications – especially in geographically distributed organizations – face the following issues:

- High storage overhead: built-in types may occupy more space than necessary, especially at billion-row scale;
- Limited precision: some DBMSs do not support nanosecond intervals, which is critical for high-frequency data (e.g., telemetry or trading);
- Complications with time zones and DST (Daylight Saving Time): automatic time conversions can introduce analysis errors;
- Inefficiency in aggregation and filtering: operations on timestamps often require extra computation, reducing performance;
- High aggregation overhead: complex computations are needed when querying events within ranges, matching timestamps, etc.

These challenges call for new approaches to storing and processing temporal data, as evidenced by the growing interest in specialized time-series databases and optimized storage formats [8]. A central aspect of optimization is the efficient storage and processing of data tied to specific timestamps. Despite the widespread use of built-in date/time types, developers regularly encounter fundamental difficulties.

Results: measured outcomes include 30–60% storage reduction, 25–40% faster queries, 35–50% faster inserts, and up to 8× throughput via batching; additional benefits are reduced network load through compression, support for high precision (up to nanoseconds), and avoidance of DST/leap-second anomalies—leading to recommendations for adopting integer timestamps in high-performance systems.

Traditional database management systems (DBMSs) provide built-in types for dates and times (DATE, TIMESTAMP, DATETIME, etc.), but they are not always optimal for high-load timeseries workloads.

## 1.2. Existing Approaches and Their Limitations

Standard data types offered by database management systems (DBMSs) often fail to account for the specific requirements of time series – such as high record density, the need for precise representation and storage of time intervals, varying granularity (from milliseconds to nanoseconds), and the need for fast, efficient search, filtering, and aggregation by time.

Moreover, common date/time storage formats frequently lead to significant performance and memory overhead. Built-in types may have limitations in temporal precision and introduce ambiguities when handling time zones and daylight-saving transitions, which further complicates temporal data processing and forces developers to build specialized solutions and algorithms for correct analysis and handling of time series.

Another major issue is that built-in types cannot fully provide the convenience and flexibility required for implementing forecasting and analytics algorithms such as smoothing, anomaly detection, and trend analysis. This forces developers to resort to additional tools or specialized systems, which can complicate the architecture and increase maintenance and operating costs.

Thus, understanding these challenges is essential for developing and refining specialized approaches and data structures designed specifically for efficient time-series processing. However, even specialized solutions do not always achieve an optimal balance among:

- Query execution speed (especially across large time ranges);
- Disk-space efficiency;
- Flexibility in data handling (e.g., support for custom time formats).

As a result, developers often turn to alternative methods for storing temporal data, including integer representations of date and time.

## 2. METHODOLOGY: INTEGER-BASED TIMESTAMP STORAGE

### 2.1. Representation Formats and Data Structure

Integer representation of timestamps is implemented using two primary formats. The first, a 32bit format (YYYYMMDD), offers a substantial storage advantage—a 40–60% reduction compared with traditional DATE types [2]. The second, a 64-bit format (YYYYMMDDHHMMSSXXXXX), provides precision down to 100 microseconds while preserving all the benefits of integer storage, as shown by TimescaleDB tests [9].

The bit layout of the 64-bit format is organized as follows (bits listed in order of significance):

- 63–56: year (most significant 2 digits);
- 55–48: month;
- 47–40: day;
- 39–32: hours;
- 31–24: minutes;
- 23–16: seconds;
- 15–0: fractional part of a second.

### 2.2. Processing and Optimization Algorithms

#### 2.2.1. Indexing Algorithms

Modified B-trees demonstrate a significant advantage when working with integer-based timestamps. In particular, PostgreSQL functional indexes show 25–40% better performance compared to standard temporal indexes, as confirmed by Leis et al. [8].

A practical implementation is as follows:

CREATE INDEX idx_events_time_int ON events (time_to_int(event_time));

#### 2.2.2. Range Search Algorithms

The optimized search process includes three key stages. At the first stage, the range boundaries are converted, which in Python can be implemented as:

```
defdate_to_int(year, month, day):    return year * 10000
+ month * 100 + day
```

At the second stage, the actual B-tree search is performed:
```
SELECT * FROM measurements
WHERE time_int BETWEEN 20230101 AND 20230131;
```

According to studies by InfluxData [10], this approach provides up to a 3× performance improvement compared to traditional formats.

### 2.2.3. Aggregation Algorithms

The binning method demonstrates particular efficiency in data aggregation:

```
SELECT
    (time_int / 10000) * 10000 AS hour_bin,   AVG(value)
FROM sensor_data
GROUP BY hour_bin;
```

Testing [6] confirmed an 18% performance advantage of this approach in analytical queries.

### 2.2.4. Batching Method for Timestamp Operations: Algorithms and Efficiency

Batching operations represent a critically important optimization for high-load time-series systems. Studies [14, 15] show that grouped operations can improve system throughput by 50–300% compared to streaming individual record processing. In the context of integer-based timestamps, batching becomes especially valuable due to modern CPU hardware optimizations for vectorized integer data processing.

For systems using integer timestamps, a modified group insert algorithm is applied, based on the research [13]:

1. Buffering:
   ```
   batch = [] for record in data_stream:
   batch.append((time_to_int(record.ts), record.value))     if len(batch) >=
   BATCH_SIZE:      execute_batch_insert(batch)       batch = []
   ```
2. Sorting (optional): uses the TimSort algorithm [14], optimized for partially ordered time sequences.
3. Batch execution: -- PostgreSQL example
   ```
   INSERT INTO metrics (ts_int, value)
   VALUES (20230101120000, 25.3),    (20230101120001,
   25.4),
        ...;
   ```

Experiments by TimescaleDB [15] show that the optimal batch size is 1,000–5,000 records for SSD-based systems and 10,000–50,000 for RAM-optimized configurations.

### 2.2.5. Timestamp Compression Algorithm

Based on the work [16], a specialized compression method has been developed:

- Delta Encoding:

```
prev_ts = batch[0][0] compressed = [prev_ts]
for ts, value in batch[1:]:
compressed.append(ts - prev_ts)     prev_ts =
ts
```

- Bit Packing: the SIMD-BP128 algorithm [17] is applied, providing 4–8× compression for sequential timestamps.

Testing on a Cassandra 4.0 cluster (16 nodes, NVMe SSD) showed the following results:

Table 1. Throughput testing results with batches

| Batch Size | Throughput (records/sec) | Latency (ms) |
|---|---|---|
| 1 (control) | 12,500 | 2.4 |
| 100 | 48,700 (+290%) | 5.1 |
| 1,000 | 112,400 (+799%) | 8.7 |
| 10,000 | 98,200 (+685%) | 102.3 |

These findings are consistent with Microsoft Azure [18], where maximum efficiency was achieved at a batch size of 2,000–5,000 records. In ClickHouse, a proprietary binary protocol is used, providing up to 1 million records/sec per node [19]. In InfluxDB, tests by InfluxData [10] demonstrate a 3–5× performance improvement with batching.
Batching operations with integer timestamps show:

- Throughput improvement up to 8×;
- Network load reduction up to 5× (due to compression);
- Optimal batch size of 1,000–5,000 records for SSD-based systems.

A promising direction is integration with hardware accelerators (GPU, FPGA) for parallel batch processing — as explored in MIT [20] and Stanford [21] projects.

### 2.2.6. Application of TAI (International Atomic Time) in Integer-Based Timestamp Storage

Standard time formats (UTC) face several issues — such as leap seconds (added periodically to synchronize with Earth's rotation), ambiguities during Daylight Saving Time (DST) transitions, and historical changes in time zones due to political or economic reasons.

TAI (International Atomic Time) solves these problems by providing linearity and continuity, while accounting for discrepancies between calendar and astronomical time. Currently, the TAI format is used in critical systems such as GPS and scientific experiments.
Conversion Algorithm:

- Obtain the current UTC time
- Apply the current TAI–UTC offset (as of 2024: 37 seconds)
- Convert to an integer-based format:
  -- Range query (ignoring leap seconds)
  SELECT * FROM scientific_data
  WHERE tai_time BETWEEN 20240101000000000000 AND 20240102000000000000;

Disadvantages of TAI:

- Requires manual updates of leap second tables;
- Debugging complexity (differences from civil time);
- Limited support in standard libraries.

Recommendations:

- Use a hybrid approach — TAI for internal calculations and UTC for user interfaces;
- Regularly update leap second data (IETF publishes official bulletins);
- For scientific systems, combine TAI with PTP (Precision Time Protocol).

TAI provides an ideal foundation for integer-based timestamp storage in systems where the following are critical:

- Linear time scale;
- High precision;
- Absence of temporal anomalies.

For most commercial systems, a hybrid model with both TAI and UTC values is recommended.

## 2.3. Implementation in Different DBMS

### 2.3.1. PostgreSQL

The recommended implementation uses GENERATED COLUMNS:

```
CREATE TABLE events (   id
SERIAL PRIMARY
KEY,event_time TIMESTAMP,
time_int BIGINT GENERATED ALWAYS AS (
EXTRACT(YEAR FROM event_time)::int * 10000000000 +
EXTRACT(MONTH FROM event_time)::int * 100000000 +
EXTRACT(DAY FROM event_time)::int * 1000000 +
EXTRACT(HOUR FROM event_time)::int * 10000 +
EXTRACT(MINUTE FROM event_time)::int * 100 +
EXTRACT(SECOND FROM event_time)::int
    ) STORED
);
```

### 2.3.2. ClickHouse

ClickHouse-Specific Optimization for Integer Timestamps:

```
CREATE TABLE metrics (ts
UInt64,    value Float64 )
ENGINE = MergeTree()
ORDER BY (ts);
```

## 2.4. Performance Optimizations

- Delta encoding: [22] showed a 60–80% reduction in data volume for high-frequency time series.
- Partitioning:
  CREATE TABLE measurements (
  time_int BIGINT,    value
  DOUBLE PRECISION )
  PARTITION BY RANGE
  (time_int);
- Vectorized processing: Using SIMD instructions delivers up to a 10× speedup [22].

## 2.5. Comparative Efficiency

Test Results (Intel Xeon 3.6 GHz, NVMe SSD, 128 GB RAM):

Table 2. Efficiency comparison.

| Operation | TIMESTAMP | Integer | Improvement |
|---|---|---|---|
| Insert 1M records | 15.2 s | 9.8 s | 35% |
| Range query | 620 ms | 380 ms | 39% |
| Daily aggregation | 1.8 s | 1.1 s | 39% |

These results confirm the findings of TimescaleDB [9] regarding the advantages of integer based storage.

## 3. PRACTICAL EXAMPLES OF INTEGER-BASED TIMESTAMP STORAGE

### 3.1. Financial Markets and High-Frequency Trading (HFT)

In the finance and fintech sector, a prime example is a high-frequency trading platform or risk management system. The task is to analyze millions of transactions per day in real-time. The problem with using string dates is that searching by a date range involves slower string comparison and less efficient indexes. The solution with an integer date is to store the trade date as a simple integer. This leads to a notable performance gain, including faster filtering of data, more efficient index trees that speed up search, and easy database partitioning for quicker query access. This leads us to:

- the trade date is stored as TradeDate INT (e.g., 20231027).
- the query becomes WHERE TradeDate BETWEEN 20231025 AND 20231027.

With this, the expected performance gain includes:

- filtering speed: 15-30% faster data selection by date, which is critical for algorithmic trading systems.
- index efficiency: Index trees (B-Tree) for integer fields are "narrower" and more balanced, speeding up search.
- partitioning: The database can be easily partitioned by TradeDate / 100 (which gives year and month). A query for a specific month will physically access only one data section instead of scanning the entire table.

## 3.2. Telecommunication Systems and CDR Processing

Within telecommunications, a common project is a billing system and call detail record analysis. The task involves processing and storing billions of records to generate invoices and analyze customer behavior. The problem is the need to aggregate this vast amount of data by days and months. The solution is to store the call date as an integer, often alongside a separate integer for the time. This results in a performance gain characterized by much faster aggregation operations for reports, substantial disk space savings when multiplied by billions of records, and fast, simple archiving and deletion of old data through partition management. Summarizing this technically, we get:

- The call start date is stored as CallDate INT. Often, Time INT is stored separately (e.g., 133455 for 13:34:55).
- To generate a monthly report, the system simply does GROUP BY CustomerID, CallDate / 100 (where CallDate / 100 drops the day, leaving YYYYMM).

With this, the expected performance gain includes:

- Aggregation Speed: GROUP BY and SUM() operations on integer fields are executed significantly faster.
- Disk Space Savings: Multiplied by billions of records, saving even 4-6 bytes per record results in saving terabytes of disk space.
- Fast Archiving Partitioning: Old data is archived by month (WHERE CallDate< 20230101). Deleting old data also becomes a simple and fast partition drop operation.

## 3.3. Industrial IoT and Equipment Monitoring

For Internet of Things and logistics, such as a transport monitoring system, the task is to handle colossal data streams from thousands of devices. The problem is the need to quickly find data for a specific device from a recent period. The solution is to split the timestamp into two integer fields: one for the event date and one for the event time. The performance gain here includes faster data insertion into pre-defined partitions, rapid querying by accessing only a single daily data partition, and a more compact data format for the resource-constrained devices themselves.

Technically basic application with each device sending a data packet every few seconds requires the timestamp to be split into two fields: EventDate INT and EventTime INT. This allows an efficient data partitioning by hours within a day. Main performance gains relate to:

- insert speed: inserting integer values into pre-created partitions is faster than inserting into a "monolithic" table with a DATETIME type.
- fast data "pruning": a query like "show the vehicle's route for October 25, 2023" is executed as an access to a single partitionEventDate = 20231025. The system doesn't sift through data from all time.
- compactness on the device: on the IoT devices themselves (with limited memory and other hardware resources), it's often simpler and cheaper to generate and transmit a packet with integer fields than to format a date-time string.

## 3.4. Use of Integer Timestamps in Industrial Projects

While companies rarely publish detailed technical blog posts when and how they have switched to integer dates, including benchmarks, there is compelling evidence and well-documented patterns from industry leaders and massive-scale systems that strongly advocate for this approach.

Some specific, known projects and companies where integer date storage is a fundamental part of their high-performance architecture are below.

Meta (Facebook): Scuba is Meta's real-time analytics database, used for monitoring thousands of internal services, A/B test results, and performance metrics. It needs to ingest billions of data points per day and allow sub-second queries over this data. Authors highlight 32-bit integerbased time storage [23] that is optimal for Scuba's goal of "blazing fast" queries. That also contributes into the efficient in-memory processing (as fixed-width columns enable rapid scans and comparisons). It also unlocks:

- simplified partitioning: data is automatically partitioned by the integer date, making old data expiration trivial.
- compact storage: combined with dictionary encoding, this minimizes memory footprint, which is critical for an in-memory database.

The core matching engines and market data feeds of exchanges like NASDAQ, NYSE, or CME Group, and the trading systems of high-frequency trading (HFT) firms tend to base on the integer databases. This is a quite secretive industry, but the requirements are public knowledge. Market Data feeds (e.g., ITCH, OUCH, FIX/FAST) often transmit timestamps as integers (nanoseconds since midnight or epoch). The order book and trade databases inside these systems are designed for nanosecond latency. The LMAX Disruptor, a high-performance interthread messaging library from a financial company, is built around the principle of avoiding object allocation and using primitive long values for everything, including timestamps, to eliminate GC pauses.

There is no room for parsing or object creation in a critical path that must execute in less than a microsecond. Storing and comparing a long integer for a timestamp is a single CPU instruction. Using a java.util.Date or LocalDateTime object would involve method calls, memory overhead, and potential garbage collection, all of which are forbidden in the hot path.

ClickHouse itself is a real-world "project" used by thousands of companies (e.g., Cloudflare, Uber, Bloomberg) for massive-scale analytics. ClickHouse has dedicated column types for storing dates and times as integers: Date, Date32, and DateTime. The documentation recommends these types for performance. Under the hood, Date is stored as a 16-bit integer (days since 1970-01-01), and DateTime is stored as a 32-bit integer (Unix timestamp).

The entire ClickHouse engine is built for speed on large scans. Using integer types allows for:

- vectorized execution: the engine can load chunks of integer dates into CPU registers and compare them in batches.
- efficient compression: integer sequences compress extremely well with algorithms like Delta+ZigZag+RLE or LZ4.
- direct use in indexes (skip indexes): integer values are ideal for creating fast, compact data skipping indexes.

User behavior analytics platforms at companies like Amazon, Booking.com, or The Trade Desk, looks into the approach. While specific internal details are proprietary, the architecture patterns are public through talks and open-source contributions. Apache Druid, a popular real-time analytics database used in many of these companies, strongly recommends using integer timestamps for its __time column. Druid's segments are partitioned based on this integer time column, and query performance is directly tied to the efficiency of time-range filters on this integer.

For queries like "show me the click-through rate for ad campaign X per hour over the last week", the ability to quickly filter the fact table by an integer time range is the foundation of performance. Storing the date as 20231027 and the hour as 0 to 23 allows for incredibly fast and simple GROUP BY and WHERE operations.

The pattern to disclosure the use of integer date storage is unequivocal: when dealing with petabyte-scale data, sub-second query requirements, or microsecond-level latency, the overhead of complex date/time objects and string parsing becomes a significant bottleneck. Replacing them with simple integers is a non-negotiable optimization employed by leading engineering team.

### 3.5. Future Scope

Standardization & Interoperability: Develop unified storage/processing standards for integer timestamps (formats, time-scale conventions, metadata for leap-second handling) to ensure cross-DBMS portability and simpler tool integration.

Hybrid Time Models: Formalize best-practice patterns for dual TAI/UTC pipelines (TAI internally for linear math; UTC at interfaces), including automated offset/table management and debugging aids.

Hardware Acceleration: Explore GPU/FPGA offload for batched inserts, range scans, and compression codecs to push beyond current 8× throughput gains in software-only paths.

Compression & Vectorization: Design next-gen integer-timestamp codecs (delta + bit-packing variants, SIMD-first implementations) tailored to real-world burstiness and multi-granularity time, with pluggable engines in major DBMSs.

Query Optimizer Awareness: Add optimizer rules and cost models that natively understand integer time binning/partitioning to improve plan choice for range filters, rollups, and timewindow joins.

Developer Ergonomics: Provide libraries, ORM mappers, and GENERATED-COLUMN templates that make integer-time adoption turnkey (conversion helpers, safe casting, lint rules), reducing the gap with native DATE/TIMESTAMP types.

Distributed Systems & Time Sync: Build reference designs for integer-time in distributed settings (CDC, replication, multi-region) with guidance on PTP/TSN synchronization and failure-mode handling around leap seconds.
Benchmarking & Guidance: Publish open, reproducible benchmarks across DBMSs and workloads (finance, telecom, IoT) to refine sizing rules for batching, partitioning, and precision selection.

When not to use INT: Codify decision checklists (small projects, heavy date-arithmetic use, ORM-centric stacks) to prevent misuse and document safe fallbacks to native temporal types.

Computer Science & Information Technology (CS & IT)                                    191## 4. CONCLUSION

The conducted analysis demonstrates that integer-based timestamp storage is an efficient methodology for modern time-series processing systems, combining high performance, precision, and resource efficiency. This approach provides significant performance improvements:

- Insert operations accelerated by 35–50%;
- Query execution time reduced by 25–40%;
- System throughput increased up to 8× through batching;
  optimized resource utilization:
- Storage volume reduced by 30–60%;
- Efficient use of CPU hardware capabilities;
- Reduced network load through compression techniques;
  enhanced accuracy and reliability:
- Nanosecond precision in mission-critical systems;
- Elimination of time anomalies (DST, leap seconds); ☐ Increased legal and audit reliability of temporal data.

The universality of this method is particularly noteworthy, confirmed by successful implementations across domains – from high-frequency trading to scientific experiments. At the same time there remains strong potential for further improvement, particularly through:

- Integration with emerging hardware architectures (GPU, FPGA);
- Application of quantum synchronization methods;
- Development of standards for distributed systems.

Thus, integer-based timestamp representation can be recommended as a modern industry standard for systems where the following are critical:

- High-performance time-series processing;
- Precision and immutability of temporal data;
- Efficient use of computational resources.

Future research in this area should focus on developing unified storage and processing standards for integer timestamps, as well as specialized hardware solutions for their optimized handling. However, sometimes it might be BETTER to use native DATE/DATETIME types:

- Small projects where readability and development simplicity are more important than micro-optimizations.
- When complex temporal operations (intervals, adding days/months) are needed, which the DBMS provides "out of the box" for its native date types.
- Projects heavily reliant on ORM (Object-Relational Mapping), where working with native date types is more idiomatic.

In modern DBMS, native DATE types are also well-optimized, but in the extreme scenarios described above, switching to INT continues to provide a tangible and sometimes critically important advantage.